\documentclass[twocolumn,showpacs,preprintnumbers,amsmath,amssymb]{revtex4}
\usepackage{graphicx}% Include figure files
\usepackage{dcolumn}% Align table columns on decimal point
\usepackage{bm}% bold math

\begin{document}

\title{Stimulated Raman backscattering of laser radiation in
deep plasma channels}
\author{Serguei Kalmykov}\email{ kalmykov@physics.utexas.edu}\author{Gennady
Shvets}
 \affiliation{The University of
Texas at Austin, The Department of Physics, Institute for Fusion
Studies, One University Station C1500, Austin, Texas 78712}
\date{\today}

\begin{abstract}
Stimulated Raman backscattering (RBS) of intense laser radiation
confined by a single-mode plasma channel with a radial variation
of plasma frequency greater than a homogeneous-plasma RBS
bandwidth is characterized by a strong transverse localization of
resonantly-driven electron plasma waves (EPW). The EPW
localization reduces the peak growth rate of RBS and increases the
amplification bandwidth. The continuum of non-bound modes of
backscattered radiation shrinks the transverse field profile in a
channel and increases the RBS growth rate. Solution of the
initial-value problem shows that an electromagnetic pulse
amplified by the RBS in the single-mode deep plasma channel has a
group velocity higher than in the case of homogeneous-plasma Raman
amplification. Implications to the design of an RBS pulse
compressor in a plasma channel are discussed.
\end{abstract}

\pacs{52.35 Mw, 52.40 Nk, 52.50 Jm}

\maketitle
%\newpage

\section{\label{Sec1}Introduction}
Stimulated Raman backscattering (RBS) of laser radiation in
plasmas~\cite{Andreev} is a parametric process in which a laser
beam (pump wave) is backscattered off the electron plasma density
fluctuations. These density perturbations are driven and amplified
by the ponderomotive beatwave of pump and scattered
electromagnetic (EM) waves. Under certain phase matching
conditions a positive feedback loop develops that results in the
onset of a temporal or spatio-temporal instability~\cite{Lindman}.
The RBS in transversely homogeneous plasmas has been extensively
studied since early 1970s~\cite{Andreev,Lindman}, when it first
came to the fore in the context of fast electron generation and
target pre-heat in laser confinement fusion. The basic treatment
of the RBS in homogeneous plasmas is now a classic that can be
found in a number of textbooks~\cite{Kruer}.

Although RBS of long laser beams has been studied for at least
three decades, the short-pulse regimes of this instability have
only recently attracted attention due to the advances in
generation and amplification of sub-picosecond multi-terawatt
laser pulses~\cite{CPA}. The RBS of such pulses in rarefied
homogeneous plasmas ($\omega_p\ll\omega_0$, where $\omega_0$ is a
fundamental frequency of laser, $\omega_p=\sqrt{4\pi e^2 n_0/m_e}$
is an electron Langmiur plasma frequency, $-|e|$ is an electron
charge, and $n_0$ is an electron plasma density) was explored in
detail in both
experiment~\cite{Darrow,french_rbs,ting96,Krushelnik,Wang,Jones}
and
theory~\cite{Tripathi,Bingham,Mora0,Pesme,Sakharov,Andreev12,Shvets0}.
The recent upsurge of interest in the RBS has been specifically
caused by theoretical discovery of the possibility to amplify and
compress ultra-short laser pulses in a plasma by backscattering a
long low-intensity counter-propagating laser
beam~\cite{Shvets3,Shvets2,Suckewer}.

Certain applications of short-pulse lasers, such as novel X-ray
source development~\cite{Burnett}, generation of high harmonics of
laser radiation~\cite{Lompre}  and laser particle
acceleration~\cite{tajima_dawson,modena} in tenuous plasmas,
benefit from a long interaction distance. In a homogeneous plasma
the region of high intensity interaction is confined to
approximately one Rayleigh diffraction length $Z_R=k_0r_0^2/2$,
where $k_0$ is a wavenumber, and $r_0$ is a radius of a focal spot
of a laser beam. Propagation over longer distances requires some
form of laser guiding. Guiding by a plasma channel is the most
promising experimental
approach~\cite{durfee95,ehrlich96,volfbeyn99,nikitin99,gaul00} for
high laser intensities. Excitation of relativistic plasma waves in
channels was
analyzed~\cite{kats_channel,gena_channel,andreev97,Li} for
particle acceleration.

The guided laser pulses are not immune to parametric
instabilities, such as forward and near-forward stimulated Raman
scattering (SRS) in
parabolic~\cite{Andreev12,Shvets5,valeo74,Andreev3,Esarey,Sprangle1,Sprangle2},
tapered~\cite{Sprangle_last}, leaky~\cite{Antonsen}, and
single-mode flat channels~\cite{Shvets1}, or large-angle SRS in
plasma-filled capillaries~\cite{McKinstrie}. Commonly, the energy
losses of a laser pulse and excessive plasma heating due to the
large-angle SRS (including RBS) are undesirable for some
applications~\cite{Kruer2}, and uncovering new physical mechanisms
that enable the RBS suppression is up to date. For some
applications, however, the RBS can be useful. E.g., the laser
pulse leading front depletion by the RBS may seed either forward
SRS or resonant modulational
instability~\cite{Sakharov,Sprangle_mod}. Novel schemes of
short-pulse amplification in a plasma~\cite{Shvets3,Shvets2} are
based on the backscattering of a long moderately intense pump
laser beam into a short counter-propagating signal pulse: the
energy of the pump is absorbed by the signal as it is amplified
and compressed. The Raman compression could be a viable path to
obtaining high-power single-cycle pulses. For transversely
homogeneous plasmas, one of the challenges of Raman amplification
is to ensure the uniformity of plasma along the interaction axis
so that the signal be downshifted from the pump by almost exactly
the plasma frequency $\omega_p$. In the present paper we discover
novel features of the RBS in plasma channels, which are favorable
for realization of Raman amplification (to anticipate, the RBS
process in a deep plasma channel can be broadband enough to make
exact resonant detuning of pump and signal unnecessary).

To enable analytic progress and to facilitate qualitative
understanding, we focus on plasma channels that support a single
confined hf EM mode, referred to as a fundamental mode of channel
(FMC). The laser pulse confined in a channel is assumed to be the
FMC. The RBS in a single-mode channel proceeds differently than in
a homogeneous plasma and can be characterized by the following
novel features: (a) reduction of the RBS peak temporal growth
rate; (b) broadening of the RBS amplification bandwidth; (c)
modification of the transverse profile of the scattered mode from
that of FMC. Depending on the parameters of channel (such as
on-axis plasma density and density depression) and pump laser
(such as frequency and intensity), those modifications can be
either more or less prominent.

It was suggested earlier~\cite{Shvets1} in the context of Raman
forward scattering (RFS) that plasma frequency variation across
the channel may significantly reduce the peak growth rate. The
effect is resulted from a strong localization of a scattering
electron plasma wave (EPW) near the channel axis. However, due to
the complicated hybrid EM nature of a relativistic EPW in a
channel~\cite{kats_channel,gena_channel,andreev97,Li}, only
approximate results were obtained in~\onlinecite{Shvets1}.
Specifically, it was assumed that the scattered radiation was in
the fundamental channel mode. For the RBS, the EM component of the
short-wavelength EPW may be neglected, enabling us to account for
the transverse profile modification of the backscattered
radiation. This is accomplished by expanding the transverse
profile of the scattered field into the channel eigenmodes, i.e.,
the FMC plus continuum of the non-bound hf EM modes. These
continuum modes of channel (CMC) do not exponentially decay
outside of the channel (as the FMC does) and exhibit a cosine-like
behavior at infinity, which makes them similar to the transverse
Fourier harmonics of scattered radiation in a homogeneous plasma.
The problem is complicated by the fact that the FMC and CMC are
not independent. Radial shear of the plasma density couples them
to each other not only creating the field structure different from
that of FMC but also affecting spectral features of the
instability. One of our goals is to evaluate the continuum-mode
contribution to the RBS growth rate. We put emphasis on the regime
of strong plasma wave localization (SPWL). Spectral features of
this regime are markedly different from those of the
homogeneous-plasma RBS. The SPWL occurs when the parameter
$\eta=(\Delta\omega_p/\omega_{p1})^2/(2\gamma_{\rm
hom}/\omega_{p1})$ is large compared to unity \{here and
elsewhere, $\gamma_{\rm hom}=\sqrt{|{\bf a_0}|^2\omega_0
\omega_{p1}/4}$ is the maximum increment of the weakly coupled RBS
in a homogeneous plasma~\cite{Andreev,Lindman},
$\omega_{p1}=\omega_p(x=0)$ is a plasma frequency on axis, $\Delta
\omega_p^2$ is the channel depth, and ${\bf a}_0=e{\bf
E}_0/(m_e\omega_0c)$ is a normalized amplitude of electric field
of the pump wave\}, which physically means that the plasma
frequency depression in a channel exceeds the RBS bandwidth in a
homogeneous plasma. In the SPWL regime, the maximum temporal
increment is shown to scale as $\gamma_{\rm hom}/\sqrt[3]{\eta}$.
Hence, plasma wave localization suppresses the instability. On the
other hand, the transverse shear of the plasma frequency results
in a broader instability bandwidth. Spectral maximum of the
backscattered light is found to be red-shifted from the pump
frequency by more than $\omega_{p1}$, and the spectrum itself
extends on the red side far beyond the frequency bandwidth of the
homogeneous-plasma RBS. The bandwidth increase is due to the
backscattering off the channel regions with higher local plasma
frequencies and, therefore, higher red-shifts of scattered light.
In the SPWL regime, the CMC contribution transversely shrinks the
fastest-growing mode of scattered radiation and concentrates the
scattered field near the channel axis. This effect is followed by
enhancement of the RBS (in the numerical examples presented in the
paper, the continuum modes add from  25\% to 50\% to the peak
value of temporal increment).

The CMC contribution to the RBS process may be neglected if the
pump amplitude is sufficiently small (or the channel is deep) to
satisfy the inequality $u_0\ll(\Delta\omega_p/
\omega_{p1})^2/\left(\omega_0/\omega_{p1} \right)^{5/4}$. This
regime is referred to as single-mode, because the scattered field
is fairly well represented solely by FMC. That only three unstable
waves (two bound EM modes and a localized EPW) participate the
single-mode RBS makes it the channel analog of the three-wave
(i.e., weakly coupled) RBS in a homogeneous plasma despite the
vast difference in spectral features. In the single-mode regime an
analytic solution of the initial-value problem has been found,
which describes the evolution of the single-mode backscattered EM
signal in the field of a single-mode pump. It is shown that the
maximum of the wave packet moves with the velocity $2c/3$, which
is higher than a group velocity of radiation $c/2$ of the weakly
coupled RBS in a homogeneous plasma. High group velocity of the
amplified pulse and its broad bandwidth produced by the transverse
shear of plasma density profitably distinguish the single-mode
Raman amplification in a plasma channel from its
homogeneous-plasma counterpart~\cite{Shvets2}.

The paper is organized as follows. In Sec.~\ref{Sec2} the
single-mode planar plasma channel is introduced, and basic
equations governing the nonlinear plasma response to the combined
pump and scattered radiation are derived. These equations are
solved in Sec.~\ref{Sec3} by the mode expansion, and the
generalized dispersion relation is derived. This dispersion
relation allows for the coupling between the FMC and CMC of
scattered radiation. In Sec.~\ref{Sec4}, the dispersion equation
is solved in the most interesting and novel regime of SPWL. The
criterium which allows to neglect the CMC contribution to the RBS
process is proposed. The spectral bandwidth and peak temporal
growth rate of the single-mode regime are derived. In
Sec.~\ref{Sec5} the initial value problem describing the linear
evolution of the EM signal is formulated and solved, and the group
velocity of the signal is evaluated. Summary of the results is
given in Sec.~\ref{Sec6}. The Appendix reviews some properties of
the associated Legendre functions with an imaginary order which
describe the continuous spectrum of the EM channel eigenmodes.

\section{\label{Sec2}Basic equations}
To begin with, we define the unperturbed state of a plasma and
laser radiation, that is, the background on which the instability
grows. In order to make analytic progress, a planar plasma channel
is chosen with the density profile given by
\begin{equation}
\label{1} \omega_{p}^2(x)=\omega_{p2}^2-\left(\Delta
\omega_p^2/2\right)\cosh^{-2}(x/\sigma).
\end{equation}
Plasma frequency in the channel varies between
$\omega_{p1}=(\omega_{p2}^2-\Delta\omega_p^2/2)^{1/2}$ in the
center and $\omega_{p2}$ at infinity. Relative plasma density
depression is
$n_e(x=0)/n_e(|x|=\infty)=(1+\Delta\bar{\omega}_p^2/2)^{-1}$,
where $\Delta\bar{\omega}_p=\Delta \omega_p/\omega_{p1}$. Through
the rest of the paper, electron density at the channel axis is
held fixed, and the channel depth is varied. The normalized hf
electric field of pump radiation with an arbitrary polarization is
\begin{equation}
\label{2} \tilde{\bf a}_0(x,z,t)={\rm Re}\left[{\bf
a}_0(x)e^{ik_0z-i\omega_0t}\right].
\end{equation}
The laser electric field in the channel is the solution of the
eigenvalue problem
\begin{equation}
\label{3} {\cal L}_0{\bf a}_0\equiv\left[-\partial^2/\partial x^2
+ k_p^2(x)\right]{\bf a}_0=\lambda_0{\bf a}_0
\end{equation}
with the boundary condition $a_0(|x|\to\infty)\to0$ [here,
$\lambda_0\equiv(\omega_0/c)^2-k_0^2$,
$k_p(x)\equiv\omega_{p}(x)/c$, and $a_0=|{\bf a_0}|$]. In
Eq.~(\ref{3}), we assume $k_{p1}\sigma\sim O (1)$,
$k_{p1}=\omega_{p1}/c$, and $a_0\ll1$ and neglect a relativistic
correction to the mass of electron oscillating in the pump field.
Thus, relativistic self-focusing of a laser beam~\cite{Sun} is
excluded. Relativistic self-guiding effects will be addressed in
future publications. We require that the eigenvalue
problem~(\ref{3}) has the unique solution decaying at
$|x|\to\infty$, which gives the transverse profile of the FMC
\begin{equation}
\label{2.1} a_0(x)=u_0\cosh^{-1}(x/\sigma)\equiv u_0\psi_0
\end{equation}
and a relation between the plasma channel depth and width
\begin{equation}
\label{3.1} \Delta\omega_p\sigma/c=2.
\end{equation}
Also, the eigenvalue equation gives the dispersion relation for
the pump field
\begin{equation}
\label{3.2}
 \lambda_0=\omega_0^2/c^2-k_0^2=k_{p2}^2-\sigma^{-2},
\end{equation}
where $k_{p2}=\omega_{p2}/c$.

The perturbed hf electric field in a plasma is
\begin{equation}
\label{4} \tilde{\bf a}(x,z,t)=\tilde{\bf a}_0 + {\rm
Re}\left[{\bf a}_s(x,z,t)e^{-i\omega_st+ik_sz}\right],
\end{equation}
where ${\bf a}_{s}=e{\bf E}_{s}/(m_e\omega_0c)$ ($|{\bf a}_{s}|\ll
a_{0}$) is a complex amplitude of the normalized electric field of
backscattered radiation, $\omega_s=\omega_0-\omega_{p1}$,
$k_s=-k_0+k_{p1}$. In the case of rarefied plasma
($\omega_p\ll\omega_0$), which is considered here, the RBS is a
resonant process~\cite{Lindman} in which only the Stokes component
of scattered radiation is involved [see Eq.~(\ref{4})]. The
amplitudes $a_{0(s)}$ are slowly varying in time and space on the
scales $\omega_0^{-1}$ and $k_0^{-1}$, respectively. We shall
consider the weakly coupled RBS~\cite{Andreev,Lindman}, whose
temporal increment is smaller than the electron Langmuir frequency
(which is valid at $u_0<\sqrt{\omega_p/\omega_0}$). Hence, the
envelope of scattered radiation is slowly varying in time  and in
the  direction of propagation $z$: $|\partial a_{s}/\partial
t|\ll\omega_{p1(2)}|a_{s}|$, $|\partial a_{s}/\partial z|\ll
k_{p1(2)}|a_{s}|$.

We neglect the ion density perturbation produced by the laser and
scattered radiation.  We consider ions to be a fixed neutralizing
positive background in the form of a channel with a density
profile given by Eq.~(\ref{1}). This assumption is adequate for
laser pulses shorter than an ion plasma period
($\tau_L\ll2\pi\omega_{pi}^{-1}$). In the opposite limit of long
pulses, various parametric instabilities have been analyzed
previously~\cite{valeo74}. Ponderomotive force due to the
interference of incident and scattered radiation excites
perturbations of electron density, $\delta {\tilde
n}_e=n_e-n_0(x)$,
\begin{equation}
\label{5} \delta {\tilde n}_e(x,z,t)={\rm Re}\left[\delta n_e(
x,z,t)e^{-i\omega_{p1}t+ik_ez}\right],
\end{equation}
where $k_e=k_0-k_s$. In a rarefied plasma, the amplitude $\delta
n_e$ of the scattering EPW  varies in space slowly on the scale
$k_0^{-1}$. Moreover, for the regime of weak coupling, this
amplitude is slowly varying on plasma temporal and spatial
periods: $|\partial \delta n_e/\partial t|\ll\omega_{p1(2)}
|\delta n_e|$, $|\partial \delta n_e/\partial z|\ll k_{p1(2)}|
\delta n_e|$.

The amplitudes of scattered EM wave and scattering EPW obey the
coupled-modes equations, which follow from the equations of
nonrelativistic hydrodynamics of electron fluid in the hf EM
field~(\ref{4}) and Maxwell's equations for the scattered
radiation,
\begin{subequations}
\label{A}
\begin{eqnarray}
 \left[\frac{\partial}{\partial t}\!-c\frac{\partial}{\partial z}
 + i\frac{c^2}{2\omega_s}\left({\cal
L}_0\!-\!\lambda_s\right)\right]a_s &\! = \!&
\omega_{p1}G_1\psi_0\nu,
\label{6}\\
\left[ \frac{\partial}{\partial t}\! - \! i\omega_{p1}
(\Delta\bar{\omega}_p/2)^2 \left(1\!-\!\psi_0^2\right)\right] \nu
&\! = \! &\omega_{p1}G_2\psi_0a_s\label{7},
 \end{eqnarray}
 \end{subequations}
where $\nu=[\omega_p(x)/\omega_{p1}]^2[\delta n_e^*/n_0(x)]$,
$G_1=u_0/(4i\bar{\omega}_s)$,
$G_2(x)=i\bar{\omega}_0^2\bar{\omega}_p^2u_0$ with
$\bar{\omega}_{0,s}=\omega_{0,s}/\omega_{p1}$,
$\bar{\omega}_p=\omega_p(x)/\omega_{p1}$,  and
$\lambda_s=(\omega_s/c)^2-k_s^2$. The difference between
$\lambda_s$ and $\lambda_0$ is eliminated from all the further
equations by the substitution
$a_s,\nu\propto\exp[icz(\lambda_0-\lambda_s)/(2\omega_s)]$. The
set~(\ref{A}) is valid under assumption that the short-wavelength
scattering plasma wave ($k\approx 2k_0\gg k_{p1(2)}$) is
predominantly electrostatic, with the EM component of the plasma
wake neglected~\cite{Shvets1}. We scale time and space to
$\omega_{p1}^{-1}$ and $k_{p1}^{-1}$, respectively, and introduce
the dimensionless variables $\bar{t}=\omega_{p1}t$,
$\left\{\bar{z},\bar{x}\right\}=k_{p1}\left\{z,x\right\}$ (all the
overbarred quantities which appear in the equations below are
normalized in this way). Eqs.~(\ref{A}) recast now as
\begin{subequations}
\label{B}
\begin{eqnarray}
 \left[\frac{\partial}{\partial\bar{ t}}  -  \frac{\partial}{\partial\bar{ z}}
\! + \! \frac{i}{2\bar{\omega}_s}\left(\bar{\cal L}_0 \! - \!
\bar{\lambda}_0\right)\right]a_s & = & \, G_1\psi_0\nu,
\label{6.1}\\
\left[\frac{\partial}{\partial\bar{t}}-i
(\Delta\bar{\omega}_p/2)^2\left(1-\psi_0^2\right)\right] \nu
&=&G_2\psi_0 a_s.\label{7.1}
\end{eqnarray}
\end{subequations}
In Sec.~\ref{Sec3}, we derive from Eqs.~(\ref{B}) the dispersion
relation of RBS in the channel~(\ref{1}).

\section{\label{Sec3}General dispersion relation}

Eqs.~(\ref{B}) can be solved by using the Fourier-Laplace
transform of the envelopes, $a_s,\nu\propto\exp (i\bar{\omega}
\bar{t}-i\bar{ k}\bar{z})$. It is convenient to introduce a new
variable $y=\tanh(x/\sigma)$, and express $\psi_0 = \sqrt{1  -
y^2}$, $\bar{\omega}_p^2 = 1 + C^2 y^2$, where
$C^2=\Delta\bar{\omega}_p^2/2=2/\bar{\sigma}^2$. Plasma occupies
the area $-1 < y < 1$. Expressing $\nu$ through $a_s$ from
Eq.~(\ref{7.1}), inserting it into Eq.~(\ref{6.1}), and using
Eqs.~(\ref{3.1}),~(\ref{3.2}), we obtain the equation for
$a_s(y,\bar{k},\bar{\omega})$,
\begin{eqnarray}
{\cal L}_{1,1}a_s = \bar{\sigma}^2\left[\frac{p^2}{1-y^2}+
\frac{G}{\bar{\omega}} {\cal G}(\bar{\omega},y)\right]a_s,
\label{8}
\end{eqnarray}
where $p^2= 2\bar{\omega}_s(\bar{\omega}+\bar{k})$,
$G=u_0^2\bar{\omega}_0^2/2$, ${\cal G}(\bar{\omega},y)=(1 + C^2
y^2) /(1-B^2y^2)$, $B^2=\Delta\bar{\omega}_p^2/(4\bar{\omega})=
\left(\bar{\sigma}^2\bar{\omega}\right)^{-1}$. The function
$p^2(\bar{k})$ contains the information about propagation of
scattered radiation. The operator in the LHS of Eq.~(\ref{8})
comes from the equation for the associated Legendre
functions~\cite{Ryzhik}
\begin{equation}
{\cal L}_{\mu,s}P_s^\mu(y) \!\equiv \! \Biggl\{\frac{d}{dy}(1 \! -
\! y^2)\frac{d}{dy} \! + \! s(s \! + \! 1) \! - \! \frac{\mu^2}{1
\! - \! y^2}\Biggr\}P_s^\mu(y) \! = \! 0 \label{9}
\end{equation}
with $s=\mu=1$. The degree $s=1$ given, the spectrum of
Eq.~(\ref{9}) consists of one discrete level with $\mu=1$ and a
continuum of modes with $\mu^2=-q^2$, $q$ real. Hence, the full
set of solutions of Eq.~(\ref{8}) is composed of the single FMC
$P_1^1(y)=\psi_0$ and the set of CMC
\begin{eqnarray}
P_1^{\pm iq}(y) & = & \frac{\tanh(x/\sigma)\mp iq}{(1\mp
iq)\Gamma(1\mp iq)}e^{\pm
iqx/\sigma}\nonumber\\
 & = & \frac{1}{\Gamma(1\mp iq)}\frac{y\mp iq}{1\mp
iq}\left(\frac{1+y}{1-y}\right)^{\pm iq/2}.\label{10}
\end{eqnarray}
At $|x|\to\infty$, the CMC~(\ref{10}) reveal the cosine-like
behavior [see Eqs.~(\ref{A2})]. Unlike the pump field given in the
form of the FMC, the scattered radiation in a channel is not
necessarily represented by FMC only. The CMC (\ref{10}) describe
the discrepancy between the true transverse field profile and the
FMC. Therefore, we express the general solution of Eq.~(\ref{8})
as a mode expansion
\begin{equation}
\label{9.1}
a_s(y)=P_1^1(y)+\int\limits_{-\infty}^{+\infty}P_1^{iq}(y)a_s(q)\,dq.
\end{equation}
Knowing that $ {\cal L}_{1,1}P_1^1 = 0$, ${\cal
L}_{1,1}P_1^{iq}(y) = -(1+q^2)(1-y^2)^{-1}P_1^{iq}(y)$, we arrive
at the equation
\begin{widetext}
\begin{equation}
\int\limits_{-\infty}^{+\infty}
P_1^{iq}(y)\frac{1+q^2}{1-y^2}a_s(q)\,dq  =  -
\bar{\sigma}^2\Biggl(\frac{p^2}{1-y^2}+  \frac{G}{\bar{\omega}}
{\cal G}(\bar{\omega},y)\Biggr) \left(P_1^1(y) +
\int\limits_{-\infty}^{+\infty}
P_1^{iq}(y)a_s(q)\,dq\right).\label{15}
\end{equation}
\end{widetext}
We multiply Eq.~(\ref{15}) by $P_1^1(y)$ and integrate it over
$y$. Having in mind that
$\int_{-1}^{1}P_1^{iq}(y)(1-y^2)^{-1/2}\,dy=0$ [formula~(7.132.1)
of Ref.~\onlinecite{Ryzhik}], we arrive at the dispersion equation
\begin{equation}
p^2+\frac{G}{2\bar{\omega}}{\tilde Q}(\bar{\omega}) = -
 \frac{G}{2\bar{\omega}}
\int\limits_{-\infty}^{+\infty}F(\bar{\omega},q)a_s(q)\,dq,\label{16}
\end{equation}
where
\begin{equation}\label{16.1}
F(\bar{\omega},iq)=\int\limits_{-1}^{1} {\cal
G}(\bar{\omega},y)(1-y^2)^{1/2}P_1^{iq}(y)\,dy,
\end{equation}
and the plasma response function reads
\begin{widetext}
\begin{equation}
{\tilde Q}(\bar{\omega})  =  \int\limits_{-1}^{1}  (1  - y^2){\cal
G}(\bar{\omega},y) \, dy =  \frac{2}{B^2} \left(1  -
\frac{2C^2}{3} + \frac{C^2}{B^2}\right)  -  \frac{(B^2  + C^2)(1
 -  B^2)}{B^5} \ln  \left(\frac{1  +  B}{1  - B}\right)
.\label{13}
\end{equation}
\end{widetext}
It has been shown previously~\cite{Shvets1} that the
function~(\ref{13}) describes a purely electrostatic response of a
plasma in the limit of a broad shallow channel, that is
$k_e\sigma\gg1$. In the case of RBS $k_e\approx 2k_0$. Thus, any
channel with $\sigma>k_0^{-1}$ is wide, and the electrostatic
description is always applicable to the plasma response in the RBS
process.

In order to find a closed-form dispersion equation we have to
determine explicitly the CMC amplitudes $a_s(q)$ in
Eq.~(\ref{16}). We multiply Eq.~(\ref{15}) by $P_1^{-iq'}(y)$ and
integrate it over $y$. Using the orthogonality
condition~(\ref{A1}) and the identity
$\Gamma(1+iq)\Gamma(1-iq)=\pi q/\sinh(\pi q)$~\cite{Ryzhik}, we
obtain the expression
\begin{eqnarray}
\lefteqn{ a_s(q)  =  -\bar{\sigma}^2\left(\frac{G}{2\bar{\omega}}
\right)\frac{q/\sinh(\pi q)}{1+(\bar{\sigma}p)^2+q^2}}\nonumber\\
& & {} \times
\Biggl[F(\bar{\omega},-iq)+\int\limits_{-\infty}^{+\infty}
F_1(\bar{\omega},q,q')a_s(q)\,dq\Biggr],\label{17}
\end{eqnarray}
where the kernel $ F_1  = \int_{-1}^{1}{\cal G}(\bar{\omega},y)
P_1^{iq}(y)P_1^{-iq'}(y)\,dy$ describes the coupling between the
modes of continuum with different indices $q$. Further, we take
into account only the coupling between the FMC and CMC, and drop
the last term in brackets in the RHS of Eq.~(\ref{17}). The
resulting dispersion relation reads
\begin{subequations}
\label{C}
\begin{eqnarray}
p^2  +  \frac{G}{2\bar{\omega}}{\tilde Q}(\bar{\omega})  =
\bar{\sigma}^2\left(\frac{G}{2\bar{\omega}}\right)^2
\Phi(\bar{\omega},p),\label{19}\\
 \Phi(\bar{\omega},p)  =
\int\limits_{-\infty}^{+\infty}
\frac{F(\bar{\omega},iq)F(\bar{\omega}, \! -iq)}{1 +
(\bar{\sigma}p)^2  +  q^2}\frac{q\,dq}{\sinh(\pi q)}.\label{20}
\end{eqnarray}
\end{subequations}
Eq.~(\ref{19}) describes the spatio-temporal evolution of initial
perturbations of field and electron density~\cite{Bers}. Real and
imaginary parts of the complex frequency $\bar{\omega}$ as a
function of a real wavenumber shift $\bar{k}$ determine the
temporal evolution of spatial Fourier harmonics of the signal, and
can be used to solve the initial-value problem. In the next
Section, the complex solution $\bar{\omega}(\bar{k})$ of the
dispersion equation~(\ref{19}) (with $\bar{k}$ real) will be found
in the limit of strong plasma wave localization (SPWL). The SPWL
is achieved when the channel depth $\Delta\bar{\omega}_p^2$ is
much larger than the instability bandwidth, i.e.
$\Delta\bar{\omega}_p^2\gg |\bar{\omega}|$, or $|B|^2\gg1$. The
unstable EPW will be found confined in the near-axis area with a
transverse extent of about $|B|^{-1}$. For this regime, the
contribution of CMC to the growth rate can be important.

\section{\label{Sec4}Solution of dispersion relation in the SPWL regime }
\subsection{\label{Subsec4.1}Temporal increment of instability}

The parameter area for the SPWL regime is prescribed by the
inequality $|B|^2>1$, or
$|\bar{\omega}|<(\Delta\bar{\omega}_p/2)^2$.  Taking the estimate
$|\bar{\omega}|\sim\bar{\gamma}_{\rm
hom}=(u_0/2)\sqrt{\bar{\omega}_0}$, we establish the limitation on
the laser amplitude from above
\begin{equation}
\label{13.4}
u_0<U_0\equiv\Delta\bar{\omega}_p^2/(2\sqrt{\bar{\omega}_0}).
\end{equation}
Eq.~(\ref{13.4}) provides the same scaling in $\bar{\omega}_0$ as
the condition of weak coupling $u_0<1/\sqrt{\bar{\omega}_0}$ for
the RBS in homogeneous plasmas~\cite{Lindman}. In the SPWL regime,
the unstable electron density perturbations are localized stronger
than the EM waves. The size of the area of localization is
$\delta\bar{x}\sim2\bar{\sigma}/|B|$.

To evaluate the RHS of the dispersion equation~(\ref{19}) in the
limit $|B|^2\gg1$, we note that the integral~(\ref{16.1}) that
determines the  kernel of Eq.~(\ref{20}) can be represented as
$F(\bar{\omega},iq)=\int_{-1}^1{\cal F}(y,iq)(1-B^2y^2)^{-1}\,dy$,
where ${\cal F}(y,iq)=(1+C^2y^2)P_1^1(y)P_1^{iq}(y)$ is
independent of $B$. At $|B|^2\gg1$, only $y$ values from the close
vicinity of channel axis, that is, $|y|<|B|^{-1}$, will contribute
to the integral. Therefore, effective coupling between the FMC and
CMC occurs near the channel axis in the area with a transverse
extent of order $|B|^{-1}$. In this case, the RHS of
Eq.~(\ref{19}) can be expanded in powers of $B^{-1}$. To evaluate
the lowest-order term of the expansion, the on-axis value of
${\cal F}(y,iq)$ is taken, i.e., ${\cal
F}(0,iq)=-iq/[(1-iq)\Gamma(1-iq)]$. Then, the integral~(\ref{20})
becomes
\begin{widetext}
\begin{equation}
\Phi\left(|B|^2 \! \gg \! 1\right) \!  \approx \! \frac{1}{\pi}
\left(\int\limits_{-1}^1\frac{dy}{1 \! - \! B^2y^2}\right)^2
\int\limits_{-\infty}^{+\infty}\frac{q^2\,dq}{\left[1 \! + \!
(\bar{\sigma} p)^2 \! + \! q^2\right]\left(1 \! + \! q^2\right)
 } \! = \!  \frac{\sqrt{1 \! + \! \left(\bar{\sigma}
p\right)^2} \! - \! 1}{(\bar{\sigma} B p)^2} \ln^2\left(\frac{1 \!
+ \! B}{1 \! - \! B}\right)\approx \pi^2\frac{1 \! - \! \sqrt{1 \!
+ \! \left(\bar{\sigma} p\right)^2}}{(\bar{\sigma} B
p)^2}.\label{24}
\end{equation}
\end{widetext}
Evaluating the plasma response function in the limit of large
$|B|$ as $\tilde{Q}(\bar{\omega})\approx i\pi/B$, we reduce
Eq.~(\ref{19}) to a relatively simple form:
\begin{equation}
\label{25} p^2 \! + \! \frac{G}{2\bar{\omega}}\frac{i\pi}{B} \! =
\! \left(\frac{G}{2\bar{\omega}}\right)^2\left(\frac{\pi}{B
p}\right)^2\left(1 \! - \! \sqrt{1 \! + \! \left(\bar{\sigma}
p\right)^2}\right).
\end{equation}
The RHS of this equation represents the contribution from the CMC
which cannot be {\it a priori} neglected. The generic SPWL regime
can be subdivided into the single- and multi-mode sub-regimes.
Below, we find the boundary between them and their spectral
features.

Under the condition
$|\bar{\sigma}p|^2\sim2\bar{\omega}_0/|B|^2<1$, the CMC
contribution may be eliminated from Eq.~(\ref{25}):
\begin{equation}
p^2+[G/(2\bar{\omega})](i\pi/B) = 0. \label{16.10}
\end{equation}
Eq.~(\ref{16.10}) describes the interaction of the pump wave (FMC)
and the {\it single} FMC of scattered light via the strongly
localized EPW. Therefore, the single-mode sub-regime admits an
analogy with the three-wave RBS in a homogeneous plasma (these
processes, however, have quite different spectral properties).
Eq.~(\ref{16.10}) yields the solution with a maximum imaginary
part,
\begin{equation}
\label{27} \bar{\omega}(\bar{k}=0)=e^{i\pi/3}
\sqrt[3]{\pi^2/(2\eta)}\,\bar{\gamma}_{\rm hom},
\end{equation}
where the parameter $\eta = \Delta\bar{\omega}_p^2/
(2\bar{\gamma}_{\rm hom})>1$ is of the order of $|B|^2$. Real part
of the solution~(\ref{27}) gives the red-shift of the spectral
maximum $\Delta\bar{\omega}=
\sqrt[3]{\pi^2/(2\eta)}(\bar{\gamma}_{\rm hom}/2)$.
Eq.~(\ref{16.10}) also predicts a blue-side limitation of the RBS
bandwidth at ${\rm Re}\,\bar{\omega}_c=-\sqrt[3]{\pi^2/\eta}
(\bar{\gamma}_{\rm hom}/2)$. From the red side (${\rm
Re}\,\bar{\omega}>0$),  Eq.~(\ref{16.10}) gives no limitation, and
the spectrum has a tail far extended in this area.

The validity condition $|B|^2>2\bar{\omega}_0\gg1$ for
Eq.~(\ref{16.10}) separates the SPWL sub-regimes. It is more
restrictive than $|B|^2>1$  and may be written as
$|\bar{\omega}|<\Delta\bar{\omega}_p^2/(8\bar{\omega}_0)$.
Substituting the solution~(\ref{27}) into the latter inequality
provides the parameter area for the single-mode sub-regime:
\begin{equation}
\label{13.5} u_0< U_1\equiv
U_0/[\sqrt{\pi}(2\bar{\omega}_0)^{3/4}]\ll U_0.
\end{equation}
The larger $u_0$, the stronger the instability is driven, and the
larger is the population of CMC excited. When $u_0$ falls within
the interval
\begin{equation}
\label{13.6} U_1< u_0< U_0,
\end{equation}
the CMC contribution is no more negligible, and the RBS process
becomes essentially multi-mode. The RHS of Eq.~(\ref{25}) cannot
be omitted then, and the dispersion equation~(\ref{25}) is solved
numerically.

\begin{figure}
\includegraphics[scale=1.0]{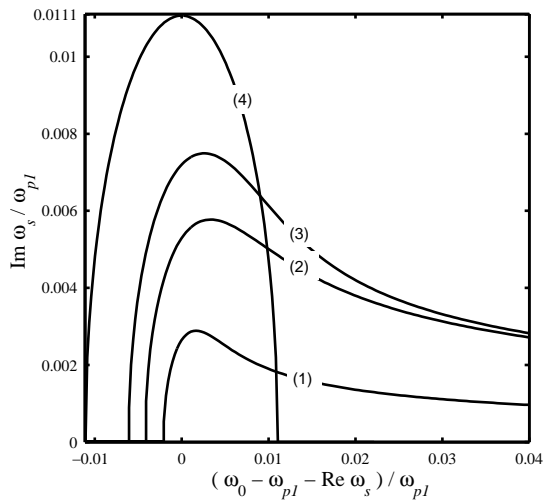}
\caption{\label{Fig1} Spectra of RBS in the regime of strong
plasma wave localization. Normalized increment is plotted against
the real part of frequency. Laser and plasma parameters are
$\bar{\omega}_0=10$ and $u_0\approx0.007$. Plasma frequency
depression in a channel is (I) $\Delta\bar{\omega}_p=2$ [curve
(1)] and (II) $\Delta\bar{\omega}_p=1/\sqrt2$ [curves (2) and
(3)]. The line (4) is a reference curve corresponding to the
homogeneous plasma ($\bar{\omega}_p=1$). Lateral variation of
$\omega_p$ red-shifts the spectral maxima and creates the tails on
the red side.  }
\end{figure}

In Fig.~\ref{Fig1}, we present an example of the dispersion curves
obtained for the RBS in both single- and multi-mode SPWL regimes.
For the fixed values of normalized laser frequency
$\bar{\omega}_0=10$ and amplitude of electric field $u_0=0.007$
the increment ${\rm Im}\,\bar{\omega}$ is found numerically versus
${\rm Re}\,\bar{\omega}$ for two different values of the channel
depth: (I) $\Delta\bar{\omega}_p=2$, or
$n_e(x=0)/n_e(|x|=\infty)=1/3$, and (II)
$\Delta\bar{\omega}_p=1/\sqrt2$, or
$n_e(x=0)/n_e(|x|=\infty)=0.8$. In Fig.~\ref{Fig1}, the solution
of full Eq.~(\ref{25}) for the set of parameters (I) is plotted
with the curve (1). The curves (2) and (3) are obtained via
numerical solution of Eq.~(\ref{25}) for the set of parameters
(II), with the CMC contribution deducted in the case of the curve
(2). The reference spectrum of RBS in a homogeneous plasma is
presented in Fig.~\ref{Fig1} by the curve (4).

Parameter sets (I) and (II) correspond to the SPWL regime as the
condition $u_0\ll U_0$ is very well satisfied for both. For the
set (I) the inequality $u_0<U_1\approx0.019$ holds, and the
single-mode regime is the case. Contribution of the CMC is
negligibly small: numerical solutions of the full [Eq.~(\ref{25})]
and single-mode [Eq.~(\ref{16.10})] dispersion equations coincide
within the thickness of the line (1). Such a good coincidence is
not the case for the parameter set (II), which corresponds to the
multi-mode regime ($u_0>U_1\approx0.0047$). Comparison between the
curves (2) and (3) demonstrates considerable enhancement of RBS
due to the CMC contribution: the CMC correction to the peak
increment amounts to about 25\%. The basic characteristics of RBS
in the multi-mode SPWL regime can be summarized as follows:
\begin{enumerate}
\item The peak growth rate is reduced if compared with the case of
homogeneous plasma. \item Contribution from the CMC enhances the
scattering process. \item The RBS bandwidth inside a channel is
significantly larger than in a homogeneous plasma. \item The
frequency spectrum experiences overall red-shift from the Stokes
frequency $\omega_0-\omega_{p1}$.
\end{enumerate}
The last two features are clear advantages of the SPWL regime for
the Raman amplification of short pulses in plasmas~\cite{Shvets2}.
Due to the broadband nature of the process, exact tuning the
signal frequency to $\omega_0-\omega_{p1}$ is not necessary to get
a considerable amplification rate in the linear regime. We can
also estimate here the RBS growth rate modification for the
parameters typical of a channel-guided laser driven accelerator.
The plasma channel created in the experiment~\cite{gaul00} was
capable of single-mode guiding of the laser pulse with a radius
8$\mu$m at the level $e^{-2}$ in intensity, which under
ansatz~(\ref{2.1}) gives $\sigma=4.65\mu$m. The electron density
at the bottom of the channel $n_1=4\times10^{18}$ cm$^{-3}$ gives
$k_{p1}\sigma=1.75$ and, according to the matching
condition~(\ref{3.1}), provides the effective normalized electron
density depression $\Delta\bar{\omega}_p^2\approx1.3$
 (we have to
mention that an actual channel shape~\cite{gaul00} is a plasma
column with a density depression at the axis surrounded by the
walls which are few laser wavelengths thick and approximately
twice the bottom density; as the walls density is much below
critical, we assume that the main contribution to the guiding
effect is made from the near-axis density profile which admits the
approximation in the form~(\ref{1}) with the effective channel
depth just calculated).
The laser wavelength $\lambda_0=0.8\mu$m
gives $\bar{\omega}_0\approx20$. To fall within the SPWL regime of
RBS the guided laser pulse must possess the intensity
$I<4\times10^{16}$ W/cm$^2$ to be in accordance with the
limitation $u_0<U_0\approx0.14$ [see Eq.~(\ref{13.4})]. For
$I\approx3\times10^{14}$ W/cm$^2$ ($u_0\approx0.012$, $\gamma_{\rm
hom}\approx0.028\omega_{p1}$) RBS proceeds in the multi-mode SPWL
regime with $\eta\approx24$. The increment formally evaluated from
Eq.~(\ref{27}) gives the peak increment
$\gamma\approx0.5\gamma_{\rm hom}$ whereas taking account of the
CMC contribution increases it to $0.8\gamma_{\rm hom}$. Therefore,
in the regime considered, 20\% reduction of the peak RBS increment
can be expected in comparison with the case of homogeneous plasma,
and the contribution from CMC to the peak increment amounts to
30\% of $\gamma_{\rm hom}$.

\subsection{\label{Subsec4.2}Transverse profile of scattered radiation in multi-mode SPWL
regime}

\begin{figure}
\includegraphics[scale=1.0]{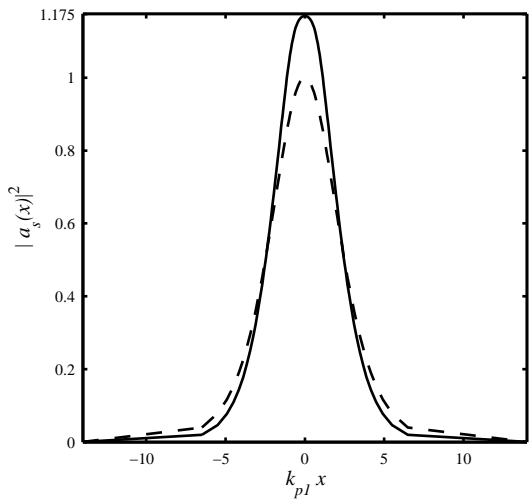}
\caption{\label{Fig2} Transverse profile of intensity for the
fastest growing spectral component of scattered radiation (solid
line) numerically evaluated from Eq.~(\ref{9.1}) for the parameter
set (II) of Fig.~\ref{Fig1}, complex frequency given by the
maximum of the curve (3). The dashed line is an intensity profile
of the bound mode. CMC contribution shrinks the transverse profile
of scattered radiation and increases the field near the channel
axis. }
\end{figure}

Results of the previous Subsection show high sensitivity of the
RBS to the transverse structure of scattered radiation in a
channel: taking account of CMC increases the value of the growth
rate. Here, we find the CMC-related correction to the transverse
profile of scattered radiation. For the parameters of the case
(II) [the curve (3) in Fig.~\ref{Fig1}] we evaluate the
integral~(\ref{9.1}) numerically for the fastest growing mode, and
present the transverse distributions of intensity
$|a_s(\bar{x})|^2$ in Fig.~\ref{Fig2}. The solid curve is given by
Eq.~(\ref{9.1}). The reference intensity profile of the FMC is
plotted with a dashed line. The coupling between the FMC and CMC
transversely compresses the scattered radiation beam. To evaluate
the effective compression numerically, we define the rms beam size
as $\sqrt{\langle\bar{x}^2\rangle}=\left(\int_{-\infty}^{+\infty}
\bar{x}^2|a_s(\bar{x})|^2\,d\bar{x}/\int_{-\infty}^{+\infty}
|a_s(\bar{x})|^2\,d\bar{x}\right)^{1/2}$. For the intensity
profiles with and without CMC shown in Fig.~\ref{Fig2} the ratio
of the rms sizes is $\sqrt{\langle\bar{x}_{\rm
FMC+CMC}^2\rangle/\langle\bar{x}_{\rm FMC}^2\rangle}\approx 0.77$.
So, the numerical example shows 23\% transverse compression of the
intensity profile reached in the multi-mode regime.
Simultaneously, for the parameters chosen, power of the scattered
radiation remains almost unchanged: difference between the
integrals $\int_{-\infty}^{+\infty} |a_s(\bar{x})|^2\,d\bar{x}$
calculated for the FMC and FMC + CMC is about 3\%. Hence, in the
multi-mode SPWL regime, coupling between FMC and CMC compresses
the scattered radiation near channel axis where the pump field has
maximum. As a consequence, increase in the peak increment occurs.
For the estimates made at the end of previous Subsection the
compression was about 30\% and intensity enhancement on the axis
about 40\%.

\section{\label{Sec5} Spatio-temporal evolution of backscattered pulse in
the single-mode SPWL regime}

\subsection{\label{Subsec5.1}Group velocity of scattered radiation}

In this Section we address the linear Raman amplification of EM
wave packet in a plasma channel. First, we evaluate the group
velocity of backscattered light, i.e., $v_g\equiv c(\partial{\rm
Re}\,\bar{\omega}/\partial\bar{k})$  at
 ${\partial{\rm Im}\,\bar{\omega}/\partial\bar{k} = 0}$. By
definition, thus calculated group velocity determines the speed of
the pulse peak. On substituting $\bar{\omega}=\Omega+i\gamma$ with
$\Omega$ and $\gamma$ real into the dispersion
relation~(\ref{16.10}), we separate real and imaginary parts of
the equation, exclude $\gamma$, and obtain the algebraic relation
which defines implicitly the dispersion function
$\Omega(\bar{k})$:
\begin{equation}
\label{29} \bar{k}^2+4\bar{k}\Omega+3\Omega^2 - \frac{\Omega^3+
2\bar{k}\Omega^2 +\bar{k}^2\Omega+b}{3\Omega+2\bar{k}}=0
\end{equation}
Differentiating Eq.~(\ref{29}) with respect to $\bar{k}$, and
plugging into the resulting equation $\bar{k}=0$ and
$\Omega(0)={\rm Re}\,\bar{\omega}$ from Eq.~(\ref{27}) (i.e., the
wave number and real part of frequency corresponding to the peak
growth rate), we find the group velocity  $v_g=-2c/3$ (the minus
sign means that the amplified pulse moves in the backward
direction).

The absolute value of the group velocity in a channel is higher
than in a homogeneous plasma, where ${v_g}_{\rm hom}=-c/2$. To get
a qualitative interpretation of this effect we return to the basic
equations~(\ref{B}). Taking the single-mode approximation for the
scattered light envelope $a_s(\bar{x},\bar{z},\bar{t})\approx
A_s(\bar{z},\bar{t})\psi_0(\bar{x})$, initial condition
$\nu(\bar{t}=0)\equiv0$, and the plasma response function
$\tilde{Q}(\bar{\omega})\approx\pi
i/B(\bar{\omega})=(2\pi/\Delta\bar{\omega}_p)\sqrt{i\bar{\omega}}$,
we reduce the set~(\ref{B}) to the single equation
$(\partial_{\bar{ t}}- \partial_{\bar{ z}})A_s \approx
u_0[\Delta\bar{\omega}_p/(16\bar{\omega}_0)]\langle
N(\bar{t})\rangle$, where the overlap integral between the
scattering electron density perturbation and transverse intensity
profile of the FMC $\langle N(t)\rangle\equiv
\int_{-\infty}^{+\infty}\psi_0^2(\delta n_e^*/n_0)\,dx = u_0
\sqrt{\pi i}(2\bar{\omega}_0/\Delta\bar{\omega}_p)^2
\int_0^{\bar{t}}A_s(\tau)/\sqrt{\bar{t}-\tau}\,d\tau$ describes
the effective plasma response. The transverse shear of the plasma
density produces an effective decay of the plasma response
expressed in terms of the convolution $\langle
N(\bar{t})\rangle\propto A_s(\bar{t})*\bar{t}^{-1/2}$ [compare
with the non-damped case of homogeneous plasma, where
$(\partial_{\bar{ t}}-
\partial_{\bar{ z}})a_s =\bar{\gamma}_{\rm hom}^2 \int_0^{\bar{t}}
a_s(\tau)\,d\tau\equiv\bar{\gamma}_{\rm hom}^2 a_s(\bar{t})*1$].
Therefore, the tail of the amplified signal experiences the growth
rate reduction according to $A_s(\bar{t})\propto \langle
N(\bar{t})\rangle\propto A_s(\bar{t})*\bar{t}^{-1/2}$, and,
consequently, the signal maximum moves closer to the signal
leading front than in a homogeneous-plasma case. This argument
qualitatively explains the increase in the group velocity of
scattered light in a channel.

\subsection{\label{Subsec5.2}Evolution of EM wave packet}

\begin{figure}
\includegraphics[scale=1.0]{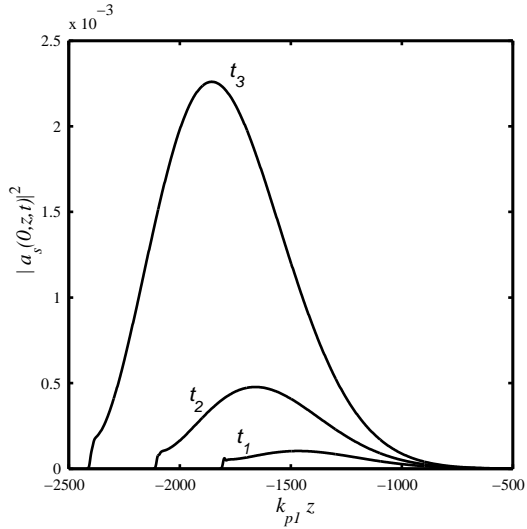}
\caption{\label{Fig3} Longitudinal on-axis profiles of intensity
of the amplified signal for the parameter set~(I) of
Fig.~\ref{Fig1}, $\bar{t}_1=1800$, $\bar{t}_2=2100$,
$\bar{t}_3=2400$. The seed signal is Gaussian, $a_{s0}=
(u_0/10)\exp\left(-\bar{z}^2/100\right)$.}
\end{figure}

We solve here the initial-value problem for the
equations~(\ref{B})
\begin{subequations}
\label{F}
\begin{eqnarray}
\label{30} a_s(\bar{x},\bar{z},\bar{t}=0) & = &
a_{s0}(\bar{z})\psi_0(\bar{x}),\\
\bar{\nu}(\bar{x},\bar{z},\bar{t}=0) & \equiv & 0\label{31}.
\end{eqnarray}
\end{subequations}
which specifies the EM wave packet initially matched with an
unperturbed plasma channel. We naturally take account of these
initial conditions by introducing a new dependent variable $
a_s^\dagger(\bar{x},\bar{z},\bar{t}) =
a_s(\bar{x},\bar{z},\bar{t})H(\bar{t})$ [where $H(\bar{t})$ is the
Heaviside step-function]. In the characteristic variables
$\theta=-\bar{z}$ and $\eta=-\bar{z}-\bar{t}$, the set~(\ref{B})
reads
\begin{eqnarray}
\left[\frac{\partial}{\partial\theta}
 +\frac{i}{2\bar{\omega}_s}(\bar{\cal
L}_0-\bar{\lambda}_0)\right]a_s^\dagger\! -\! G_1\psi_0\nu &\! =
\!& a_{s0}(\!-\theta) \psi_0\delta(\theta\!-\!\eta),
\nonumber\\
\left[ \frac{\partial}{\partial\eta} \! - \!
i\frac{\Delta\bar{\omega}_p^2}{4i}\left(1\!-\!\psi_0^2\right)\right]
\nu \! + \! G_2\psi_0a_s^\dagger &\! =\! & 0,\nonumber
 \end{eqnarray}
where $\delta(t)$ is the Dirac delta-function. We apply the
Fourier transform with respect to the co-moving ``spatial''
variable $\eta$, assuming $ \delta(\theta-\eta) =
(2\pi)^{-1}\int_{-\infty}^{+\infty} \exp[ik(\eta-\theta)]\,dk$,
and then exclude the Fourier image $\nu(\bar{x},\theta,k)$.
Multiplying the resulting equation for the Fourier image
$a_s^\dagger(\bar{x},\theta,k)$ by $\psi_0(\bar{x})$ and then
integrating over $\bar{x}$ we obtain the averaged equation
\begin{eqnarray}
\lefteqn{ \frac{1}{\left\langle
\psi_0^2\right\rangle}\Biggl\{\left\langle \psi_0 \frac{\partial
a_s^\dagger}{\partial\theta}\right\rangle \! + \!
\frac{i}{2\bar{\omega}_s} \Biggl[\left\langle
\psi_0\left(\bar{\cal L}_0 \! - \! \bar{\lambda}_0\right)
a_s^\dagger\right\rangle }\label{32}\\
& & {} -  \!  G\left\langle \frac{\psi_0^3\left[1 \! + \!
C^2\left(1 \! - \! \psi_0^2\right)\right]}{k\left[1 \! -  \!
B_k^2\left(1 \! - \!
\psi_0^2\right)\right]}a_s^\dagger\right\rangle\Biggr]\Biggr\} \!
= \! a_{s0}( \! - \theta)e^{-ik\theta},\nonumber
\end{eqnarray}
where $B_k^2\equiv -\Delta\bar{\omega}_p^2/(4k)$. In a generic
case, transverse profile of the signal $a_s^\dagger(\bar{x})$ is a
superposition of the FMC and CMC. In this Section, we consider the
single-mode SPWL regime of Raman amplification and, for all
instants of time, take the wave packet in the form of FMC [hence,
$(\bar{\cal L}_0-\bar{\lambda}_0)a_s^\dagger=0$]. We substitute
$a_s^\dagger(\bar{x},\theta,k)\equiv A(\theta,k)\psi_0(\bar{x})$
into Eq.~(\ref{32}) and get finally
\begin{equation}
\label{33} \frac{\partial A}{\partial\theta}  +
\bar{\gamma}^2_{\rm hom}\left( \frac{Q(k)}{2ik}\right)A=
a_{s0}(-\theta)e^{-ik\theta}.
\end{equation}
Here, the plasma response function [compare to Eq.~(\ref{13})]
reads
\begin{equation}
\label{34}
Q(k)=\int\limits_{-1}^1\frac{1+C^2y^2}{1-B_k^2y^2}(1-y^2)\,dy.
\end{equation}
In the SPWL regime, relatively large wave number shifts,
$|k|\gg(\Delta\bar{\omega}_p/2)^2$ (that is, $|B_k|^2\gg1$),
contribute to the signal evolution. Hence, the main contribution
to the integral~(\ref{34}) is made by the integrand values in the
vicinity of $y=0$ (or $|y|\ll|B_k|^{-1}$). Approximating the
integrand as $\left(1-B_k^2y^2\right)^{-1}$ we find the
approximate response function $Q(k)  \approx -
2\pi\sqrt{k}/\Delta\bar{\omega}_p$ which allows to present
Eq.~(\ref{33}) in the form
\begin{equation}
\label{36} \left(\partial /\partial\theta  + i{\cal
F}k^{-1/2}\right)A= a_{s0}(-\theta)e^{-ik\theta},
\end{equation}
where ${\cal F}=\pi\bar{\gamma}_{\rm hom}\sqrt{\bar{\gamma}_{\rm
hom}/(2\eta)}$. Solution of Eq.~(\ref{36}) reads
\begin{equation}
\label{37} A(\theta,k)  =  \int\limits_0^\infty a_{s0}(\theta_1 -
\! \theta) e^{  -  ik(\theta-\theta_1)-i{\cal
F}\theta_1/\sqrt{k}}\,d\theta_1.
\end{equation}
Inverting the Fourier transform~(\ref{37}) and returning to the
lab-frame variables, we find the longitudinal evolution of the
signal, $ A(\bar{z},\bar{t})  =   \int_0^\infty a_{s0}(\theta_1 +
\bar{z}){\cal D}(\theta_1  - t,\theta_1)\,d\theta_1$, where the
the Green function of RBS in the single-mode SPWL regime, $
 {\cal D}(\theta_1 \! - \! t,\theta_1) \! =  \! (2\pi)^{-1}
\int_{-\infty}^{+\infty} \exp[ik(\theta_1 \! - \! t) \! - \!
i{\cal F}\theta_1/\sqrt{k}]dk$, can be found explicitly in terms
of a generalized hypergeometric function
${}_1F_3$~\cite{Wolfram3}:
\begin{eqnarray}
\lefteqn{{\cal D}(\theta_1 \! - \! t,\theta_1)   \! = \! \delta(t
\! - \! \theta_1) \! + \!  i H(t \! - \! \theta_1)\left({\cal F}
\theta_1\right)^2} \nonumber\\&  & {} \times
{}_1F_3\left(\frac{1}{2};1,\frac32,\frac32;i({\cal F}\theta_1)^2(t
\! - \! \theta_1)/4\right).\label{40}
\end{eqnarray}
Solution of the initial-value problem expressed explicitly in
terms of the Green function~(\ref{40}) reads
\begin{widetext}
\begin{equation}
\label{39}
a_s^\dagger(\bar{x},\bar{z},\bar{t})=\psi_0(\bar{x})\left\{
a_{s0}(\bar{t}+\bar{z}) + i{\cal F}^2\int\limits_0^t
{}_1F_3\left(\frac{1}{2};1,\frac32,\frac32;i({\cal
F}\theta_1)^2(t-\theta_1)/4\right)
a_{s0}(\theta_1+\bar{z})\theta_1^2 \,d\theta_1 \right\}.
\end{equation}
\end{widetext}
In Fig.~\ref{Fig3} we present the evolution of initially Gaussian
signal $a_{s0}=(u_0/10)\exp\left(-\bar{z}^2/100\right)$ for the
parameter set (I) of Fig.~1. Snapshots of the intensity profile
are presented for three consequent instants of time. The front of
the amplified signal moves with a speed of light in the negative
$z$ direction, whereas its maximum moves with a group velocity
$v_g=-2c/3$, as predicted in the previous Subsection. The maximum
of the amplified signal grows exponentially in time with an
increment twice as given by~(\ref{27}). Therefore, the solution of
initial-value problem confirms the predictions of the dispersion
analysis of Sec.~\ref{Sec4}.

\section{\label{Sec6}Conclusion}
We have investigated the RBS of laser radiation in the regime of
strong plasma wave localization (SPWL) in a plane plasma channel,
which can support only one trapped EM mode (the single-mode
channel~\cite{Shvets1}). For the SPWL regime, transverse variation
of the plasma frequency exceeds the RBS bandwidth calculated for a
homogeneous plasma with an on-axis electron density, and the
scattering plasma wave is localized stronger than a driving
beatwave of pump and scattered radiation. In this case, the
transverse profile of the scattered radiation is a superposition
of the fundamental mode of a channel (FMC) and a continuum of
non-bound modes of channel (CMC).  Depending on the plasma and
laser parameters, the RBS with SPWL can proceed either in a
single- or multi-mode regime. In the single-mode case, excitation
of the CMC is suppressed almost completely, and only the FMC of
scattered radiation is involved in the process. This allows for a
physical analogy between the single-mode SPWL regime of RBS in a
plasma channel and a three-wave RBS in a homogeneous plasma. In
the multi-mode SPWL regime, the CMC play an essential role. The
CMC contribution transversely shrinks the scattered radiation
beam, and increases the growth rate of the instability. Spectral
features of both sub-regimes are qualitatively similar. The
temporal growth rate is slower, and amplification bandwidth is
greater than in the case of homogeneous plasma, and the frequency
spectrum experiences an overall red-shift. The group velocity of
scattered radiation, $v_g\approx-2c/3$, is increased versus its
homogeneous-plasma value, ${v_g}_{\rm hom}=-c/2$. All these
features are the consequences of the transverse shear of the
electron plasma density.  The broadband nature of RBS in the SPWL
regime and relatively high group velocity of scattered radiation
are the features profitable for the Raman amplification of short
pulses in deep plasma channels.

\begin{acknowledgments}
This work is made under financial support of the Office of High
Energy Physics of the US Department of Energy, Grant
DE-FG02-03ER41228.
\end{acknowledgments}

\appendix

\section{\label{AppA}Properties of associated Legendre functions with imaginary order}
The associated Legendre functions $P_1^{\pm
iq}(y)$~\cite{Wolfram}, where $q$ is a positive real number, are
the solutions~(\ref{10}) of Eq.~(\ref{9}) with $s=1$ and $\mu =
\pm iq$. The orthogonality condition for $P_1^{\pm iq}(y)$ reads
\begin{equation}
\frac{1}{2\pi}\int\limits_{-1}^{1} \!
P_1^{iq}(y)P_1^{-iq'}(y)\frac{dy}{1 \! - \! y^2} \! = \!
\frac{\sinh(\pi q)}{\pi q}\delta(q' \! - \! q),\label{A1}
\end{equation}
which is evaluated using the substitution $y=\tanh(x/\sigma)$,
formulas (3.982), (3.987.1), and (3.981.6)~\cite{Ryzhik} and the
Dirac delta-function property
$\delta(q'-q)=(2\pi)^{-1}\int_{-\infty}^{+\infty}e^{ix(q-q')}\,dx$.
The functions $P_1^{\pm iq}(x)$ do not vanish at
$|x/\sigma|\to\infty$ and have the asymptotic
\begin{subequations}
\label{A2}
\begin{eqnarray}
P_1^{iq}\left(\frac{x}{\sigma}\to+\infty\right) & \sim & \frac{e^{
iqx/\sigma}}{\Gamma(1- iq)},\\
P_1^{iq}\left(\frac{x}{\sigma}\to-\infty\right) & \sim &
\left(\frac{iq+1}{iq-1}\right)\frac{e^{- iq|x|/\sigma}}{\Gamma(1-
iq)}.
\end{eqnarray}
\end{subequations}
The cosine function can be constructed of $P_1^{\pm iq}(x)$ by
summing up the even complex conjugate solutions ${\cal
P}_1^{iq}(x)=[P_1^{iq}(x)+P_1^{iq}(-x)]/2$ and ${\cal
P}_1^{-iq}(x)=[P_1^{-iq}(x)+P_1^{-iq}(-x)]/2$:
\[
\cos(qx/\sigma) = \left(\frac{\pi}{2}\right)\frac{(1+q^2)[{\cal
P}_1^{iq}(x) + {\cal P}_1^{-iq}(x)]}{ {\rm
Im}\,\Gamma(1+iq)+q\,{\rm Re}\,\Gamma(1+iq)}.
\]

%\newpage

%\newpage

%\centerline{Figure captions} \hangindent=1.4cm \noindent Fig.~1.

 %\hangindent=1.4cm \noindent Fig.~2.

%\hangindent=1.4cm \noindent Fig.~3.

\end{document}